\documentclass[aps,prc,showpacs,twocolumn,floatfix,superscriptaddress]{revtex4}

\usepackage{graphicx}

\def\gtorder{\mathrel{\raise.3ex\hbox{$>$}\mkern-14mu
             \lower0.6ex\hbox{$\sim$}}}
\def\ltorder{\mathrel{\raise.3ex\hbox{$<$}\mkern-14mu
             \lower0.6ex\hbox{$\sim$}}}

\begin{document}

\title{Low-$Q$ scaling, duality, and the EMC effect}

\author{J. Arrington}
\affiliation{Argonne National Laboratory, Argonne Illinois 60439, USA}
\author{R. Ent}
\affiliation{Thomas Jefferson National Accelerator Facility, Newport News Virginia 23606, USA}
\author{C. E. Keppel}
\affiliation{Thomas Jefferson National Accelerator Facility, Newport News Virginia 23606, USA}
\affiliation{Hampton University, Hampton VA 23668, USA}
\author{J. Mammei}
\altaffiliation{present address: Virginia Tech, Blacksburg, Virginia 24061, USA} 
\affiliation{Juniata College, Huntingdon Pennsylvania 16652, USA}
\author{I. Niculescu}
\affiliation{James Madison University, Harrisonburg, Virginia 22807, USA}

\date{\today}

\begin{abstract}

High energy lepton scattering has been the primary tool for mapping
out the quark distributions of nucleons and nuclei.  Data on the proton and
deuteron have shown that there is a fundamental connection between the low and
high energy regimes, referred to as quark-hadron duality. We present the
results of similar studies to more carefully examine scaling, duality, and in
particular the EMC effect in nuclei.  We extract nuclear modifications to the
structure function in the resonance region, and for the first time demonstrate
that nuclear effects in the resonance region are identical to those measured
in deep inelastic scattering.  With the improved precision of the data at
large $x$, we for the first time observe that the large-$x$ crossover point
appears to occur at lower $x$ values in carbon than in iron or gold.

\end{abstract}

\maketitle

\section{Introduction}

Extensive measurements of inclusive lepton-nucleus scattering have been
performed in deep inelastic scattering (DIS) kinematics. In DIS kinematics,
where both the four-momentum transfer, $Q$, and the energy transfer, $\nu$,
are sufficiently large, the extracted structure function exhibits scaling,
i.e. is independent of $Q^2$ except for the well understood logarithmic QCD
scaling violations.  In this region, the structure function is interpreted as
an incoherent sum of quark distribution functions, describing the motion of
the quarks within the target.

Such measurements have unambiguously shown that the nuclear structure functions
deviate from the proton and neutron structure functions. Such modifications,
termed the EMC effect after the first experiment to observe
them~\cite{aubert83}, demonstrate that the nuclear quark distribution function
is not just the sum of the proton and neutron quark distributions. Within two
years of the first observation, hundreds of papers were published on the topic.
After 20 years of experimental and theoretical investigation, the effect still
remains a mystery.  For detailed reviews of the data and models of the EMC
effect, see Refs.~\cite{arneodo94,geesaman95}

Existing measurements of the EMC effect indicate little $Q^2$ dependence, and
an $A$ dependence in the magnitude, but not the overall form, of the structure
function modification in nuclei. The nature of the modifications in nuclei
depends primarily on Bjorken-$x$, $x=Q^2/2M\nu$, where in the parton model $x$ is
interpreted as the momentum fraction of the struck quark, and the nuclear
effects are divided into four distinct regions. In the shadowing region, $x <
0.1$, the structure function is decreased in nuclei relative to the
expectation for free nucleons. In the anti-shadowing region, $0.1<x<0.3$, the
structure function shows a small nuclear enhancement. For $0.3<x<0.7$, referred
to as the EMC effect region, the nuclear structure function shows significant
depletion.  Finally, there is a dramatic enhancement as $x$ increases further,
resulting from the increased Fermi motion of the nucleons in heavier nuclei.

Explanations of the EMC effect are hampered by the lack of a single description
that can account for the nuclear dependence of the quark distributions in
all of these kinematic regimes. Here, we will limit ourselves to $x>0.3$, the
region where valence quarks dominate. 
Even in this limited region, there is not a single explanation that can
completely account for the observed nuclear structure function modifications.
If the nuclear structure function in this region is expressed as a convolution
of proton and neutron structure functions, there are two alternative
approaches used to describe the observed medium effect: (1) incorporating
nuclear physics effects that modify the energy-momentum behavior of the bound
proton with respect to the free proton, or (2) incorporating changes to the
internal structure of the bound proton.  It has been argued, most recently
in~\cite{smith02}, that the binding of nucleons alone can not explain the EMC
effect.  In addition, several attempts to explain the EMC effect in terms of
explicit mesonic components appear to be ruled out due to limits set
by Drell-Yan measurements~\cite{alde90}. Hence, the EMC effect may be best
described in terms of modifications to the internal structure of the nucleon
when in the nuclear environment.

We note that while the EMC effect has
been mapped out over a large range of $x$, $Q^2$, and $A$, information is still
rather limited in some regions.  There are limited data on light nuclei
($A<9$), and almost no data in the DIS regime at extremely large $x$, where 
the quark distributions in nuclei are enhanced 
due to the effects of binding and Fermi motion.  Since binding and Fermi motion
impact the EMC ratios for all $x$ values, it is important to be
able to constrain these effects in a region where other, more exotic,
explanations are not expected to contribute.  It should be possible to learn
more about the EMC effect at large $x$ by taking advantage of the extended
scaling of structure functions in nuclei~\cite{filippone92, arrington01}.  In
this paper, we attempt to quantify the deviations from perturbative scaling at
large $x$, with the goal of improving measurements of the structure functions
and the EMC ratios at large $x$.

\section{Scaling of the nuclear structure function}

Inspired by a recent series of electron scattering experiments in Hall C at
Jefferson Lab, we revisit the issues of scaling in nuclear structure functions
and the EMC effect. The Hall C data are at lower invariant mass $W$, $W^2 =
M_p^2 + 2M_p\nu (1-x)$, and therefore higher $x$, than data thus far used to
investigate the EMC effect. Most notably, these new data are in the resonance
region, $W^2 < 4$~GeV$^2$. In the DIS region, $W^2 > 4$~GeV$^2$, the $Q^2$
dependence of the structure functions is predicted by perturbative QCD (pQCD),
while additional scaling violations, target mass corrections and higher twist effects, occur at
lower $Q^2$ and $W^2$ values. Thus, data in the resonance region would not naively be
expected to manifest the same EMC effect as data in the deep inelastic scaling
regime. The effect of the nuclear medium on resonance excitations seems
non-trivial, and may involve much more than just the modification of quark
distributions observed in DIS scattering from nuclei.

However, while resonance production may show different effects from the
nuclear environment, there are also indications that there is a deeper
connection between inclusive scattering in the resonance region and in the DIS
limit. This connection has been a subject of interest for nearly three decades
since quark-hadron duality ideas, which successfully described hadron-hadron
scattering, were first extended to electroproduction. In the latter, Bloom and
Gilman~\cite{bloom71} showed that it was possible to equate the proton
resonance region structure function $F_2(\nu, Q^2)$ at low $Q^2$
to the DIS structure function $F_2(x)$ in the high-$Q^2$ scaling regime, where
$F_2$ is simply the incoherent sum of the quark distribution functions. For
electron-proton scattering, the resonance structure functions have been
demonstrated to be equivalent on average to the DIS scaling strength for all
of the spin averaged structure functions ($F_1$, $F_2$,
$F_L$)~\cite{niculescu00b,liang05}, and for some spin dependent ones
($A_1$)~\cite{airapetian02} (for a review of duality measurements,
see~\cite{melnitchouk05}).

The goal of this paper is to quantify quark-hadron duality in nuclear
structure functions and to determine to what extent this can be utilized to
access poorly understood kinematic regimes. While the measurements of duality
from hydrogen indicate that the resonance structure function are on average
equivalent to the DIS structure functions, it has been observed that in
nuclei, this averaging is performed by the Fermi motion of the nucleons, and
so the resonance region structure functions yield the DIS limit without any
additional averaging~\cite{filippone92,arrington01}.

\begin{figure}
\includegraphics[height=10.0cm,width=8.2cm]{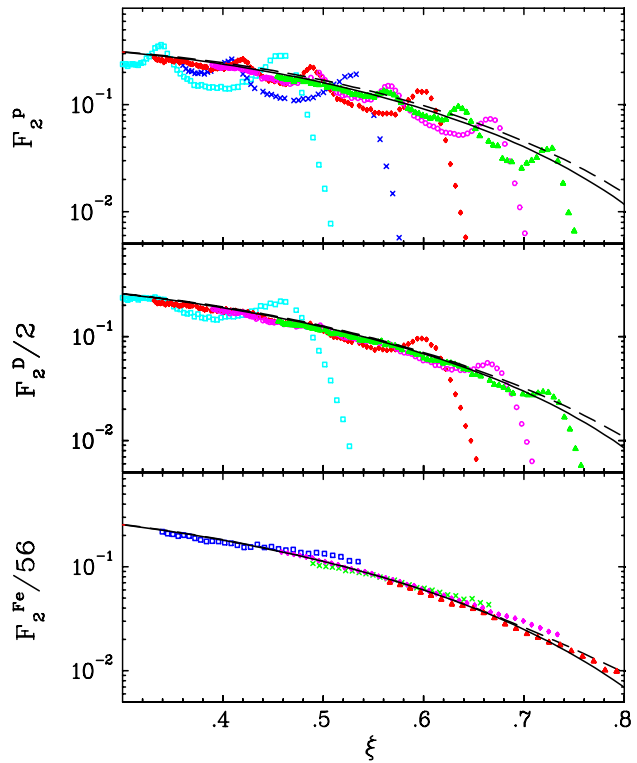}
\protect{\caption{(Color online) The $F_2$ structure function per nucleon vs
$\xi$ for hydrogen (top), deuterium (middle), and iron(bottom).  For the
hydrogen and deuterium data (0.8 $< Q^2 <$ 3.3 GeV$^2$), the elastic
(quasielastic) data have been removed.  For the iron data ($Q^2 <$ 5.0
GeV$^2$), a cut of $W^2 > 1.2$ GeV$^2$ is applied to remove the quasielastic
peak. The curves are the MRST~\cite{martin02} (solid) and NMC~\cite{arneodo95}
(dashed) parameterizations of the structure functions at $Q^2=4$~GeV$^2$, with
a parameterization of the EMC effect~\cite{gomez94} applied to produce the
curve for iron.\label{fig_h_d_fe}}}
\end{figure}

Figure~\ref{fig_h_d_fe} shows the structure functions for
hydrogen~\cite{niculescu00b}, deuterium~\cite{niculescu00a}, and
iron~\cite{arrington01}, compared to structure functions from
MRST~\cite{martin02} and NMC~\cite{arneodo95} parameterizations.  Each set of
symbols represents data in a different $Q^2$ range, with the highest $Q^2$
curves covering the highest $\xi$ values. Note that the data are plotted as a
function of the Nachtmann variable, $\xi = 2x/(1+\sqrt{1+4M^2x^2/Q^2})$,
rather than $x$. In the limit of large $Q^2$, $\xi \rightarrow x$, and so
$\xi$ can also be used to represent the quark momentum in the Bjorken limit. 
At finite $Q^2$, the use of $\xi$ reduces scaling violations related to target
mass corrections~\cite{georgi76}. The difference between $\xi$ and $x$ is
often ignored in high energy scattering or at low $x$, but cannot be ignored
at large $x$ or low $Q^2$.  The goal is to examine $\xi$-scaling to look for
any significant scaling violations beyond the known effects of perturbative
evolution and target mass corrections.
Examining the scaling in terms of $\xi$ instead of $x$ is only an approximate
way of applying target mass corrections, but it is a reasonable approximation
to a more exact correction~\cite{georgi76} in the case of the proton, and the
appropriate prescription for target mass corrections in nuclei is not as well
defined.

The transition from scaling on average in the proton to true scaling for
nuclei is clearly visible. There is significant resonance structure visible in
hydrogen, but on average the structure function reproduces the scaling curve
to better than 2\% globally and 5\% locally around each resonance for $Q^2 >
1$ GeV$^2$~\cite{niculescu00b}.  For deuterium, Fermi motion and other medium
effects broaden the resonances to the point where only the $\Delta$ resonance
has a clear peak, and the data at higher $W^2$ values, while still in the
resonance region, is indistinguishable from the scaling curve except at the
lowest $Q^2$ values.  For the iron data, taken at somewhat higher $Q^2$
values, even the $\Delta$ is no longer prominent, and deviations from pQCD
predictions are small, and limited to the tail of the quasielastic peak.

\begin{figure}
\includegraphics[height=6.0cm,width=8.2cm]{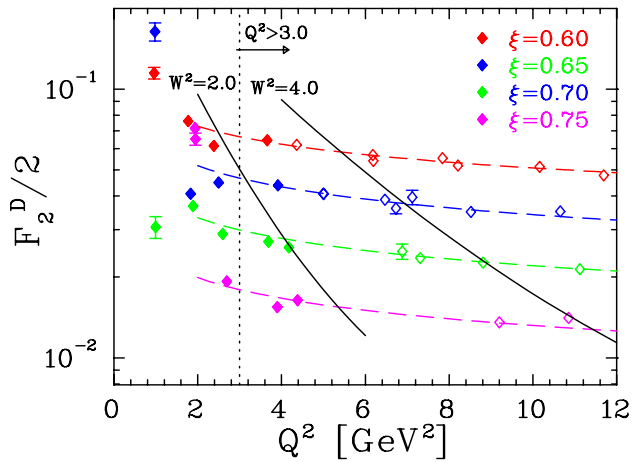}
\protect{\caption{(Color online) $F_2$ structure function per nucleon vs.
$Q^2$ for deuterium at fixed values of $\xi$.  Dashed lines show a logarithmic
$Q^2$ dependence, with the value of $d\ln{F_2}/d\ln{Q^2}$ determined at each
$\xi$ value from SLAC data at high $Q^2$ (up to 20 GeV$^2$). The solid lines
denote $W^2$=2.0 and 4.0 GeV$^2$. The combined statistical and systematic
uncertainties are shown.  The hollow symbols are data from SLAC~\cite{dasu94},
while the solid symbols are from Jefferson Lab~\cite{arrington01}.
\label{fig_xiscale}}}
\end{figure}

We can study the quality of scaling in the resonance region more directly by
examining the $Q^2$ dependence of the structure function at fixed $\xi$.
Figure~\ref{fig_xiscale} shows the $Q^2$ dependence of the structure function
for several values of $\xi$.  Above $W^2=4$~GeV$^2$, the data are in the DIS
region and the $Q^2$ dependence is consistent with the logarithmic $Q^2$
dependence from QCD evolution (dashed lines).  Even at lower $W^2$, where the
data are in the resonance region, scaling violations are small. Above
$Q^2=3$~GeV$^2$, the data deviate from the logarithmic $Q^2$ dependence by
$\ltorder 10\%$, even down to $W^2=2$~GeV$^2$.  Data on heavier nuclei show a
similar extended scaling in the resonance region, as seen in Fig.~3 of
Ref.~\cite{arrington01}, although the largest scaling violations (in the
vicinity of the quasielastic peak) are smaller, due to the increased Fermi
smearing.

Analyses of duality for the proton~\cite{armstrong01} and for
nuclei~\cite{ricco98, niculescu05} show that the moments of the structure
function, $M_n = \int x^{n-2} F_2(x,Q^2) dx$ is the $n$th moment, follow
perturbative QCD evolution down to $Q^2 \approx 2$ GeV$^2$ for the proton and
to even lower values, $Q^2 \ltorder 1$ GeV$^2$, for nuclei. The fact that the
moments follow the perturbative behavior is consistent with the observation
that the structure function in Figs.~\ref{fig_h_d_fe} and \ref{fig_xiscale}
are, on average, in agreement with the perturbative structure function.

The data indicate relatively small deviations from pQCD for $Q^2>3$~GeV$^2$ at
all values of $\xi$ measured. These deviations decrease as $Q^2$ increases,
making the nuclear structure functions at large $\xi$ consistent with the
perturbative dependence even at values of $W^2$ well below the typically DIS
limit. The limited kinematics coverage of the existing data make it difficult
to precisely map out deviations from perturbative evolution.  There is a large
gap in $Q^2$ between the JLab data shown here and the SLAC measurements at
large $Q^2$. The situation will be improved by the recently completed
measurements from JLab experiments E03-103 and E00-116~\cite{e03103,e00116},
which will provide more complete $\xi$ coverage over a wide range in $Q^2$.  In
the meantime, Fig.~\ref{fig_xiscale} indicates that for $Q^2 \gtorder
3$--4~GeV$^2$, one can relax the usual DIS requirement that $W^2 > 4$~GeV$^2$,
and the structure functions measured in the resonance region will still
provide a good approximation to the DIS structure functions.

Even with the uncertainties arising from possible higher twist contributions,
data at large $\xi$ can significantly improve our knowledge of the high-$\xi$
nuclear structure functions. There is very little DIS data for $\xi \gtorder
0.8$, and no existing facility has the combination of energy and luminosity
necessary to make precise measurements of in this regime, which requires $Q^2
> 15$(36)~GeV$^2$ for $\xi > 0.8$(0.9). If we can set reasonable limits on
scaling violations at fixed $\xi$ due to possible higher twist contributions,
we can provide useful data in this region where the few existing measurements
of the EMC effect have 10--20\% uncertainties.

\section{Structure function ratios}

Because of the difficulty in making precise measurements in the DIS region
at large $\xi$, existing measurements of the EMC effect in this region are
very poor. At large $\xi$, the EMC effect should be dominated by binding
effects and Fermi motion.  Constraining these effects will allow a better
separation of these ``conventional'' nuclear effects, which are important at
all $\xi$ values, from more exotic effects that have been used to explain the
EMC effect at lower $\xi$.

We can examine the EMC effect in the resonance region using recent
measurements~\cite{arrington01} of inclusive scattering from deuterium,
carbon, iron, and gold. For these data, we take the cross section ratio of
iron to deuterium \textit{in the resonance region} for the highest $Q^2$
measured ($Q^2 \sim 4$ GeV$^2$), requiring $W^2>1.2$ GeV$^2$ to exclude the
region very close to the quasielastic peak.

There are small differences between the analyses of the SLAC and JLab
data which had to be addressed to make a precise comparison. First, the SLAC
and BCDMS ratios were extracted as a function of $x$ rather than $\xi$.  Because
the conversion from $x$ to $\xi$ depends on $Q^2$, we can only compare ratios
extracted at fixed $Q^2$ values. Thus, for E139 we use the ``coarse-binned''
ratios, evaluated at fixed $Q^2$, rather than ``fine'' $x$ binning, which were
averaged over the full $Q^2$ range of the experiment. Coulomb corrections were
applied in the analysis of the JLab data~\cite{arrington99}, but not the SLAC
data. The SLAC data shown here include Coulomb corrections, determined by
applying an offset to the incoming and outgoing electron energy at the
reaction vertex~\cite{arrington99}, due to the Coulomb field of the nucleus.
The correction factor is $<$0.5\% for carbon, and (1.5--2.5)\% for gold.  The
JLab and SLAC ratios are corrected for neutron excess, assuming
$\sigma_n/\sigma_p = (1-0.8\xi)$.

\begin{figure}
\includegraphics[height=10.0cm,width=8.2cm]{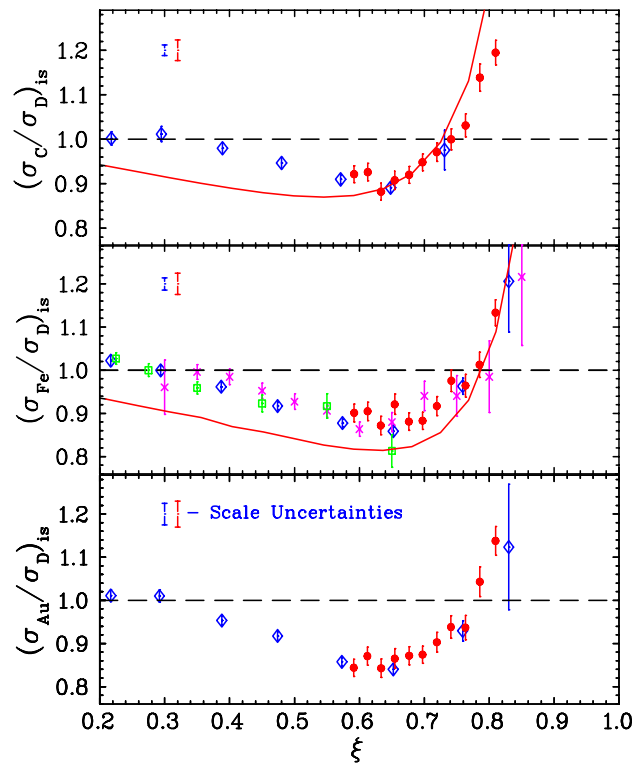}
\caption{(Color online) Ratio of nuclear to deuterium cross section per
nucleon, corrected for neutron excess. The solid circles are Jefferson lab
data taken in the resonance region ($1.2 < W^2 < 3.0$ GeV$^2$, $Q^2 \approx 4$
GeV$^2$).  The hollow diamonds are SLAC E139 data, the crosses are the
SLAC E87 data, and the hollow squares are BCDMS data, all in the DIS region. 
The scale uncertainties for the SLAC (left) and JLab (right) data are shown in
the figure.  The curves show an updated version~\cite{liuti_priv} of the
calculations from Ref.~\cite{gross92}. \label{fig_emc_all}}
\end{figure}

Figure~\ref{fig_emc_all} shows the
cross section ratio of heavy nuclei to deuterium for the previous SLAC
E139~\cite{gomez94}, E87~\cite{bodek83} and BCDMS~\cite{benvenuti87} DIS
measurements, and for the JLab E89-008~\cite{arrington99, arrington01} data in
the resonance region.  The size and $\xi$ dependence of nuclear modifications
in the JLab data agrees with the higher $Q^2$, $W^2$ data for all targets. 
Table~\ref{tab:emc} shows the ratios extracted from the JLab data.

The agreement of the resonance region data with the DIS measurement of the EMC
effect, which directly measures the modification of quark distributions in
nuclei, is quite striking.  There is no \textit{a priori} reason to expect
that the nuclear effects in resonance production would be similar to the
effects in scattering from quarks. However, it can be viewed as a natural
consequence of the quantitative success of quark-hadron
duality~\cite{niculescu00b, melnitchouk05}. As seen in Fig.~\ref{fig_h_d_fe},
the structure functions for nuclei show little deviation from pQCD, except in
the region of the quasielastic peak (and $\Delta$ resonance at low $Q^2$). As
$Q^2$ increases, the deviations from pQCD decrease as quasielastic scattering
contributes a smaller fraction of the cross section.  In retrospect, given the
lack of significant higher twist contributions, combined with the fact that
any $A$-independent scaling violations will cancel in the ratio, it is perhaps
not surprising that the resonance EMC ratios are in agreement with the DIS
measurements.

While it is difficult to precisely quantify the higher twist contributions
with the present data, we can estimate their effect by looking at low $W^2$
and $Q^2$, where the higher twist contributions are much larger. At $Q^2
\approx 2$~GeV$^2$ and $W^2 \approx M_\Delta^2$, the scaling violations (beyond
target mass corrections) for deuterium are as large as 50\%, as seen in
Fig.~\ref{fig_h_d_fe}.  However, if one takes the iron and deuterium data from
Ref.~\cite{arrington01}, \textit{averages} the structure function over the
$\Delta$ region and then forms the EMC ratio, the result differs from the
ratio in the DIS region by less than 10\%.  The decrease in the effect of
higher twist contributions is a  combination of the fact that the contribution
are reduced when averaged over an adequate region in $W^2$~\cite{niculescu00b,
melnitchouk05}, and cancellation between the higher twist contributions in
deuterium and iron.  The same procedure yields 2--3\% deviations from the EMC
ratio if one looks in the region of the $S_{11}$ or $P_{15}$ resonances, where
the scaling violations in the individual structure functions are smaller to
begin with.

For the ratios in Fig.~\ref{fig_emc_all}, we expect even smaller higher
twist effects because the data is nearly a factor of two higher in $Q^2$ and
is above the $\Delta$ except for the very highest $\xi$ points.  At higher
$Q^2$, the higher twist contributions in the individual structure functions
become smaller, while averaging over the resonance region becomes less
important as the resonances become less prominent.  Thus, we expect that
higher twist contributions for these data will be smaller than the the 2--3\%
effect ($<$10\% near the $\Delta$) observed on the EMC ratio at $Q^2\approx
2$~GeV$^2$.  If so, the higher twist corrections will be small or negligible
compared to the large statistical uncertainty in previous measurements, and
this data can be used to improve our knowledge of the EMC effect at large
$\xi$.

\begin{table}
\begin{center}
\begin{tabular}{|c|c|c|c|c|}
\hline
 $\xi$ & $W^2$ & ($\sigma_C/\sigma_D$)$_{is}$ &($\sigma_{Fe}/\sigma_D$)$_{is}$ & ($\sigma_{Au}/\sigma_D$)$_{is}$  \\
 & GeV$^2$ & & & \\ \hline
  ~0.592~  &  2.86  &  ~0.921$\pm$0.012~  & ~0.901$\pm$0.013~ & ~0.844$\pm$0.013~ \\
  0.613  &  2.70  &  0.926$\pm$0.013  & 0.905$\pm$0.014 & 0.871$\pm$0.015 \\
  0.633  &  2.55  &  0.882$\pm$0.013  & 0.872$\pm$0.015 & 0.843$\pm$0.016 \\ \hline
  0.654  &  2.39  &  0.908$\pm$0.014  & 0.921$\pm$0.017 & 0.865$\pm$0.017 \\
  0.676  &  2.22  &  0.920$\pm$0.012  & 0.881$\pm$0.012 & 0.872$\pm$0.014 \\
  0.697  &  2.07  &  0.948$\pm$0.011  & 0.883$\pm$0.010 & 0.874$\pm$0.013 \\ \hline
  0.719  &  1.91  &  0.971$\pm$0.013  & 0.917$\pm$0.012 & 0.903$\pm$0.016 \\
  0.741  &  1.75  &  1.000$\pm$0.016  & 0.976$\pm$0.015 & 0.938$\pm$0.020 \\
  0.763  &  1.59  &  1.031$\pm$0.019  & 0.964$\pm$0.017 & 0.937$\pm$0.023 \\ \hline
  0.786  &  1.43  &  1.139$\pm$0.022  & 1.013$\pm$0.020 & 1.043$\pm$0.028 \\
  0.810  &  1.26  &  1.195$\pm$0.014  & 1.133$\pm$0.015 & 1.138$\pm$0.024 \\ \hline
\end{tabular}
\caption{Isoscalar EMC ratios for carbon, iron, and gold in the resonance
region extracted from the data of Ref.~\cite{arrington01}.  $W^2$ is
calculated using the nucleon mass, rather than the nuclear mass.}
\label{tab:emc}
\end{center}
\end{table}

\section{The EMC effect at large $x$ ($\xi$)}

A careful examination of the crossover point at large $\xi$, where the ratio
$(\sigma_A/\sigma_D)_{is}$ becomes larger than unity, reveals that this
appears to occurs at larger $\xi$ for heavy nuclei than for light nuclei. This
behavior is consistent with the SLAC data, but the large-$\xi$ coverage of the
previous measurements was insufficient to make a clear statement about the
crossover point.  This observation
contradicts the argument that the dramatic enhancement at large $\xi$ is simply
due to increased Fermi motion in heavy nuclei relative to deuterium.  In this
simple picture, the slight increase in Fermi motion as one goes to heavier
nuclei would lead to an earlier onset of this enhancement.  While the Coulomb
corrections and neutron excess corrections do affect the $A$ dependence, the
uncertainties on these corrections are not large enough to explain the
observed differences between carbon and heavier nuclei.

Within the convolution formula of proton and neutron structure functions, this
crossover comes about due to counteracting contributions at large $\xi$ of the
average nucleon binding energy and average kinetic energy \cite{kulagin94},
and is hardly expected to change for $A >$ 10.  More detailed calculations
have been done to determine the effect of binding and Fermi motion, but except
for calculations of $^{3,4}$He, most of these calculations were performed for
a Fermi gas model or for infinite nuclear matter, with the density varied to
approximate the finite nuclei. These calculations do not show any significant
change in the $\xi$ dependence for heavy ($A>10$) nuclei.  Because of the lack
of precise data at large $x$, especially with respect to the $A$-dependence,
realistic models of the nuclear structure were generally not considered to be
necessary.

The effect we observe was predicted in a calculation by Gross and Liuti
\cite{gross92} using a manifestly covariant form of the convolution formula.
The most significant difference was an additional binding correction due to
the explicit dependence of the bound nucleon structure function on the
momentum of the bound nucleon. Their calculation predicts a shift in the
high-$\xi$ crossover point between carbon and iron, somewhat larger than is
observed in the data.  Another calculation including $A$-dependent nuclear
spectral functions~\cite{marco96} also gave an $A$-dependent crossover point
at large $\xi$.  However, this calculation had the crossover point moving to
\textit{lower} $\xi$ values for heavier nuclei.

Other models of the EMC effect have looked at physics beyond Fermi motion and
binding.  As with the binding models, they generally did not attempt to
reproduce the detailed $A$-dependence, and instead evaluated the EMC effect
for nuclear matter as a function of density.  Most of these models
were designed to describe the excess strength at lower $\xi$, and in general
they do not significantly impact the structure function at large $\xi$.
Thus, the addition of improved EMC ratio measurements at large $\xi$ and the
observation of an $A$ dependence to the high-$\xi$ behavior is most important
in constraining the portion of the EMC effect that is related to binding.
One can see from the calculations shown in Fig.~\ref{fig_emc_all} that the
effects of binding and Fermi motion are important over the entire $\xi$
region, and not just at the largest $\xi$ values.  These conventional
nuclear effects must be well constrained to establish a reliable baseline
before one can isolate any additional nuclear modification at lower $x$ values
that might require a more exotic explanation.
Improved data at large $\xi$ and for a variety of nuclei should allow for
tests of the prescriptions chosen for binding and Fermi motion, and thus
provide a more reliable baseline for models of the EMC effect.

The modified $x$ dependence in carbon also appears to contradict the
conclusions of a recent effective field theory calculation of the EMC effect
that predicts factorization of the $A$ and $x$ dependence, and thus the
universality of the $x$ dependence~\cite{chen05}.  The change in the high-$x$
crossover in the present data is small, but a recent measurement of the
EMC effect for $^3$He and $^4$He~\cite{e03103} will provide a more sensitive
test of the universality of the $x$ dependence.

\section{Conclusions}

This analysis provides not only an increase in the $\xi$ range of the
measurements of nuclear structure functions, but also the first observation of
an $A$ dependence of the high-$\xi$ behavior of the EMC effect.  Measurements
utilizing higher energy beams will extend measurements of the EMC effect to
even larger $\xi$ values. Based on the results shown here, the uncertainties
on extracting the EMC effect at large $\xi$ due to higher twist contributions
will be small, if not negligible, compared to the uncertainties of existing
data. A recent measurement at Jefferson Lab~\cite{e03103} will extend
measurements of the EMC effect to larger $\xi$ values and to light nuclei,
where few-body calculations can be performed with significantly smaller
uncertainties coming from uncertainties in the nuclear structure. The
calculations of Ref.~\cite{gross92} predict that the high-$\xi$ crossover
point in carbon occurs at lower $\xi$ than in heavier nuclei, as was observed
in the data.  However, they predict a crossover at much \textit{larger} $\xi$
for $^4$He.  Similarly, Ref.~\cite{burov99} also predicts a different
high-$\xi$ behavior in $^4$He than in heavy nuclei, and in addition predicts a
significant difference between $^3$He and $^4$He.

Similar investigations of duality and scaling in polarized and separated
structure functions are underway~\cite{e02109, e01006, e01012}. If duality in
these processes is quantitatively as successful as in this case, this will
have a similar impact on our ability to measure high-$\xi$ polarized structure
functions.

In conclusion, we present the first extraction of the nuclear dependence of
the inclusive structure function in the resonance region.  The data are in
agreement with previous measurements of the nuclear dependence of the quark
distributions in DIS scattering measurements of the EMC effect. This
surprising result can be understood in terms of quark-hadron duality, where
the structure function in the resonance regime is shown to have the same
perturbative QCD behavior as in the DIS regime. These data expand the $\xi$ and
$Q^2$ range of such measurements, and provide the first new measurement of the
EMC effect for a decade. They also indicate the possibility for dramatic
improvements in both the $\xi$ and $A$ range in future measurements, using the
higher beam energies currently available at Jefferson Lab.

\begin{acknowledgments}

This work was supported in part by DOE Grants W-31-109-ENG-38 and 
DE-FG02-95ER40901, and NSF Grant 0099540.  We thank
Simonetta Liuti for providing updated calculations, and Dave Gaskell
for useful discussions.

\end{acknowledgments}

\bibliography{nuclear_dual}

\end{document}